 \documentclass[12pt]{elsarticle}

\usepackage{hyperref}

\makeatletter
\providecommand{\doi}[1]{%
  \begingroup
    \let\bibinfo\@secondoftwo
    \urlstyle{rm}%
    \href{http://dx.doi.org/#1}{%
      doi:\discretionary{}{}{}%
      \nolinkurl{#1}%
    }%
  \endgroup
}
\makeatother

\usepackage{amssymb}
\usepackage{xcolor}
\usepackage{url}
\usepackage{lineno}

\journal{NIM A}

\begin{document}
\newcommand{\insertref}{{\color{red}[REF]}}

\begin{frontmatter}



\title{Measurement of the non-linearity in the $\gamma$-ray response of the GAGG:Ce inorganic scintillator}


\author[inst1,inst2]{Riccardo~Campana}
\author[inst1,inst3]{Chiara~Evola}
\author[inst1]{Claudio~Labanti}
\author[inst4]{Lisa~Ferro}
\author[inst4]{Miguel~Moita}
\author[inst1,inst2]{Enrico~Virgilli}
\author[inst1]{Ezequiel~J.~Marchesini}
\author[inst1,inst4]{Filippo~Frontera}
\author[inst4]{Piero~Rosati}

\affiliation[inst1]{organization={INAF/OAS},
            addressline={Via Gobetti 101}, 
            city={Bologna},
            postcode={40129}, 
            country={Italy}}

\affiliation[inst2]{organization={INFN, Sezione di Bologna},
            addressline={Viale Berti Pichat 6/2}, 
            city={Bologna},
            postcode={40127}, 
            country={Italy}}

\affiliation[inst3]{organization={University of Bologna, Department of Physics and Astronomy},
            addressline={Via Gobetti 93}, 
            city={Bologna},
            postcode={40129}, 
            country={Italy}}

\affiliation[inst4]{organization={University of Ferrara, Department of Physics},
            addressline={Via Saragat 1}, 
            city={Ferrara},
            postcode={44124}, 
            country={Italy}}

\begin{abstract}
A characteristic of every inorganic scintillator crystal is its light yield, i.e., the amount of emitted scintillation photons per unit of energy deposited in the crystal. Light yield is known to be usually non-linear with energy, which impacts the spectroscopic properties of the scintillator.
Cerium-doped gadolinium-aluminium-gallium garnet (GAGG:Ce) is a recently developed scintillator with several interesting properties, which make it very promising for space-based $\gamma$-ray detectors, such as in the HERMES nanosatellite mission. In this paper we report an accurate measurement of the GAGG:Ce non-linearity in the 20--662~keV $\gamma$-ray energy interval, using a setup composed of three samples of GAGG:Ce crystals read out by Silicon Drift Detectors (SDDs).
\end{abstract}




\end{frontmatter}


\section{Introduction}
\label{sec:intro}
The HERMES (\emph{High Energy Rapid Modular Ensemble of Satellites}, \cite{fiore20NR}) pathfinder mission is based on a constellation of nano-satellites flying in a low-Earth orbit (LEO). The main aim is the study of high-energy transient events in the Universe, such as \emph{Gamma Ray Bursts} (GRB), which are sudden, energetic and bright flashes of X and $\gamma$-ray photons resulting from the collapse of massive stars and from the coalescence of compact objects, such as binary neutron star systems. 

The mission is centered on six 3U CubeSat platforms, hosting a detector based on a hybrid Silicon Drift Detector (SDD) and scintillator system \cite{evangelista22NR}, sensitive to X-rays and gamma-rays, is funded both by the Italian Space Agency (ASI) and by the European Union Horizon 2020 Research and Innovation Programme.
A seventh, identical HERMES payload is onboard the Space Industry Responsive Intelligent Thermal (SpIRIT) 6U CubeSat \cite{auchettl22}. The launch of both missions is foreseen in late-2023/mid-2024.

The HERMES detection system exploits the so-called \emph{siswich} principle \cite{fuschino20NRb}, in which a SDD plays the dual role of direct X-ray detector and of photosensor for scintillation light readout. 
The SDD \cite{gattirehak84}, thanks to a series of opportunely designed cathodes implanted on the silicon surface, shape the internal electric field in order to drift the charge generated by ionising radiation to a small collecting anode. The decoupling of the ionisation-sensitive area and the charge collecting area offers several advantages, for example in terms of electronic noise. In the siswich design, low
energy photons (about 2--60 keV) are absorbed directly in the silicon bulk, while higher energy photons (from about 20~keV up to about 2~MeV) interact with the scintillator crystal, producing optical photons which are in turn detected by the same SDD. The discrimination between the two types of events in the HERMES detector design (the so-called ``X'' and ``S'' modes)  is made by reading out a crystal with two independent SDD cells: if a trigger occurs on only one SDD, it is recognized as a X-ray event, while a simultaneous trigger on both SDDs is associated with a scintillation event.

\begin{figure}
    \centering
    \includegraphics[width=\textwidth]{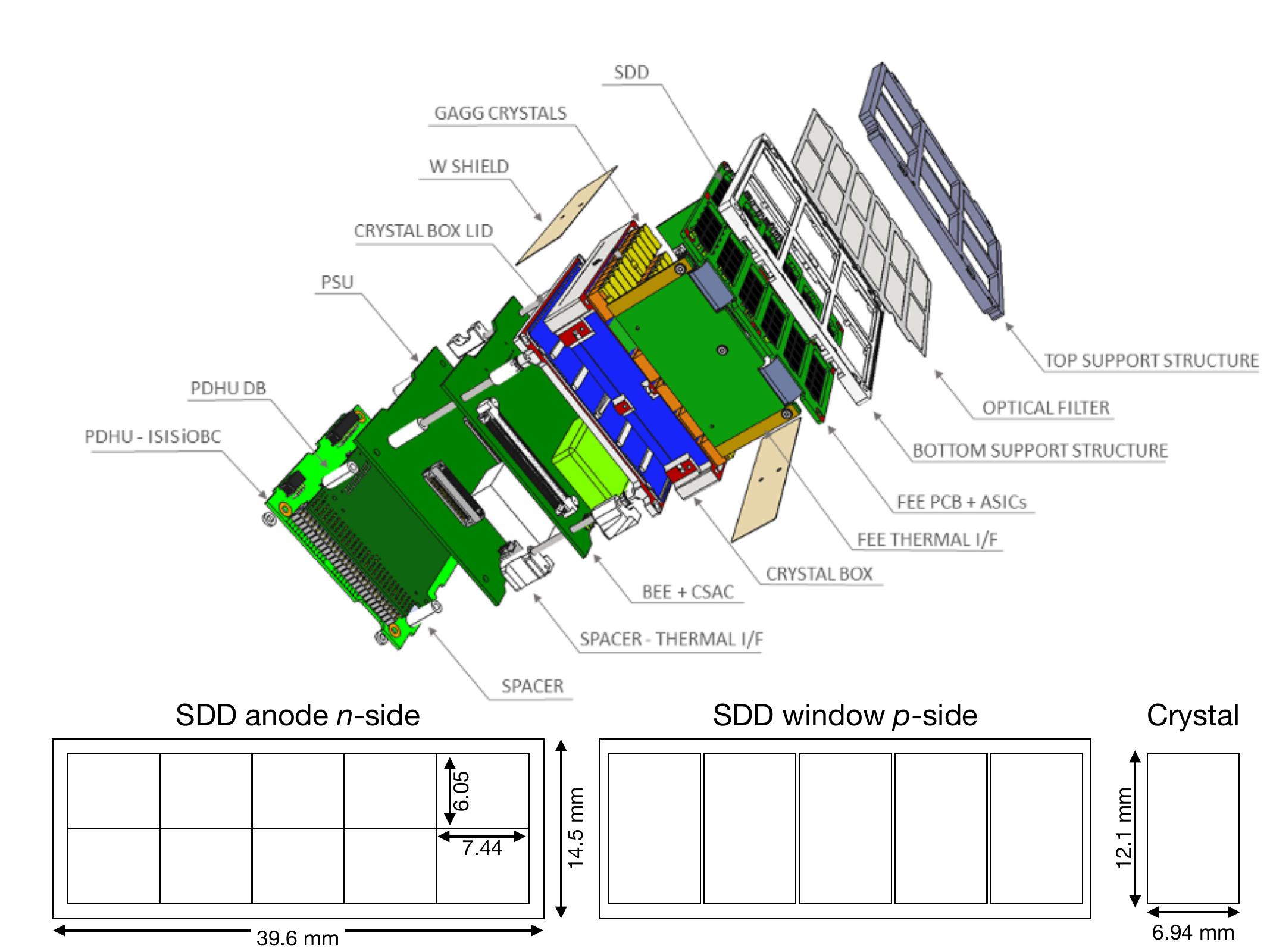}
    \caption{Exploded view of the HERMES Pathfinder scientific payload, showing the location of the 12 SDD arrays (10 independent cells each) and of the 60 GAGG:Ce crystals. The payload volume envelope is within one CubeSat unit (1U, about $10 \times 10 \times 10$ cm$^3$), while the whole satellite (including, e.g., batteries, avionics, etc.) is 3U. In the lower panel, a drawing of the HERMES SDDs is shown, with the relevant dimensions.}
    \label{fig:hermes_payload}
\end{figure}

The HERMES detector (Figure~\ref{fig:hermes_payload}) hosts 120 SDD channels each with $\sim$45 mm$^2$ sensitive area (distributed over twelve 5$\times$2 monolithic matrices) coupled with 60 cerium-doped ga\-do\-li\-nium-alum\-inium-gallium garnet (Gd$_3$Al$_2$Ga$_3$O$_{12}$:Ce or GAGG:Ce) scintillators, each 15~mm thick with a $6.94 \times 12.10$~mm$^2$ section. 
The crystal dimensions were chosen to maximise the effective area given the geometrical, mechanical and physical constraints imposed by the CubeSat platform and the SDD design.

GAGG:Ce \cite{kamada12} is a rather recently developed scintillator, with several characteristics which make it very attractive for a high-energy detector. It is luminous (light yield $\sim$50,000~photons/MeV), fast (characteristic scintillation time $\sim$90~ns), non-hygroscopic, and dense (6.63~g/cm$^3$). It has an high average atomic number ($Z \sim 54.4$) and thus a significant radiation stopping power, its scintillation light wavelength is well matched with typical silicon photosensor quantum efficiencies ($\sim$520~nm) and has no background induced by self-radioactivity.
Moreover, preliminary results \cite{sakano14,yoneyama18,dilillo22} have shown that this crystal shows a significant hardness to radiation, a particularly interesting property for applications in space satellite missions.

A characteristic of almost every scintillator is the so-called \emph{non-linearity} or \emph{non-proportionality} in its response \cite{valentine98,moses08}, that is, the luminous efficiency or light output (i.e., the amount of light emitted per unit of absorbed energy) is not constant with respect to the energy, depending on several factors. One consequence is the degradation of the spectroscopic performance of a detector based on such a material.
GAGG:Ce has not yet been extensively investigated in this regard. A few works \cite{sibczynski15,sibczynski18,furuno21,gao22} have shown a light output decrease of a few percent for photon energies below a few hundreds of keVs, and the presence of a discontinuity due to the $K$-edge of the gadolinium (at 50.23 keV), but a detailed description of the non-linearity phenomenon in this material is still lacking.

However, in order to properly calibrate the response of the HERMES detector in a large range of $\gamma$-ray energies \citep{campana22}, the deviation of the light output from the reference value (usually assumed at the 662~keV energy from $^{137}$Cs) should be carefully evaluated.

In this paper we discuss an accurate photon non-linearity measurement over a large energy interval ($\sim$20--662 keV). The response and the light output of samples representative of the HERMES flight crystals were measured using both monochromatic X-ray beams and radioactive calibration sources.

The paper is structured as follows. In Section~\ref{sec:nonlin} we summarize the physical origins of the scintillation non-linearity, while in Section~\ref{sec:setup} we describe our setup. The measurements and the results are discussed in Sections~\ref{sec:meas} and \ref{sec:res}, respectively, while in Section~\ref{sec:disc} we draw our conclusions.

\section{Scintillator non-linearity}
\label{sec:nonlin}
As mentioned above, the scintillator non-linearity 
arises from different physical processes. An incident $\gamma$-ray can interact in several ways with the scintillator crystal, transferring its energy to electrons through several concurrent processes (e.g., Compton scattering, photoelectric absorption, radiative and non-radiative/Auger atomic de-excitation, etc.), which could depend also on the crystal size and geometry. In other words, for a given incident $\gamma$-ray energy there is a probability distribution for the resulting electron energies. In turn, the scintillation response to a given electron energy can be a rather complex process. This variation in the event-to-event response is a significant contributing factor to the overall scintillator energy resolution. 
While in the literature the terms \emph{proportionality} and \emph{linearity} are often used interchangeably when referring to a linear relation $L=aE+b$ between the energy $E$ deposited in a crystal and the amount $L$ of scintillation photons produced (e.g., \cite{zerby61}, see also \cite{valentine98b, knoll10}), the former term, strictly speaking, should refer to the case in which $b=0$.

The precise behavior of the light output dependence on photon energy is different for each scintillator material, within three main categories \cite{swiderski12}: ``excess'' scintillators (e.g., NaI:Tl) in which there is a significant increase in the light yield around 100~keV; ``deficit'' scintillators (e.g., LSO:Ce) with a large drop in light yield below 100~keV; and ``quasi-proportional'' scintillators (e.g., LaBr$_3$:Ce) where the light yield deficit is less than a few percent down to 10--20~keV \cite{moses12}. In general, the response to photons is smooth, except in proximity of the binding energies of the innermost electron shells ($K$ and $L$-shells), in which the marked discontinuity in the energy distribution of electrons, following a photoelectric absorption, results in a discontinuity in the light output at the value corresponding to the atomic $K$ or $L$-edge. The ``jump'' can be as high as of 4\%--5\% \cite{ferreira04}.
As it will be shown in the following, the GAGG:Ce scintillator can be considered belonging to the quasi-proportional scintillator category.

\section{Experimental setup}
\label{sec:setup}
\subsection{Detector}
The experimental setup is based on a $1\times1$~cm$^2$ SDD, designed by INFN-Trieste and produced by FBK (Trento, Italy) in the framework of the ReDSoX collaboration\footnote{\url{http://www.redsox.iasfbo.inaf.it}}. The SDD $p$-side optical contact window is encapsulated in an epoxy layer for mechanical robustness. The SDD is mounted on a custom-made printed circuit board (PCB), which distributes the provided necessary power supplies and HV biases (about $-$180 V for the outermost SDD cathode). 

The SDD anode is connected through a wire bonding to a commercial charge-sensitive preamplifier (CSA), the XGLab CUBE\footnote{\url{https://www.xglab.it/products/cube/}}, which is hosted in a $\sim 1\times 1$~mm$^2$ die placed as close as possible to the SDD to minimize stray capacitance. A suitable control board (XGLab XGL-CBB-1CH) provides the CSA ramp resetting and bias voltages.

The preamplifier output is then fed to a commercial multichannel (Amptek DP5\footnote{\url{https://www.amptek.com/products/digital-pulse-processors/dp5-digital-pulse-processor-and-mca}}), which removes the preamplifier ramp, digitizes the signal with a fast ADC and performs a real-time digital shaping with a trapezoidal filter\footnote{For details, see, e.g., the Amptek application notes available at \url{https://www.amptek.com/-/media/ametekamptek/documents/resources/dpp_theory.pdf} and references therein.}, thus directly acquiring and storing the event spectrum. 
The trapezoidal filter shaping time was optimized to reduce the overall noise: for our data, an optimal rise time of  8~$\mu$s was used.
Longer and shorter shaping times will give rise to different absolute values of the signal amplitude, due to the ballistic deficit effect, but the overall behaviour of the light output vs. energy, which is the quantity of interest here, does not change. The trapezoidal filter flat-top duration was fixed at 0.5~$\mu$s.

Other configuration parameters for the DP5 MCA are summarized in Table~\ref{t:mca}.

\begin{table}[htbp]
    \centering
    \begin{tabular}{c|c}
        Parameter & Value\\ \hline
        Peaking time &  8~$\mu$s \\
        Flat-top duration &  0.5~$\mu$s \\
        Total gain factor &  $\times$2.494 \\
        Reset lockout duration &  20~$\mu$s \\
        Total number of ADC channels & 1024 \\
    \end{tabular}
    \caption{Some DP5 MCA setting parameters used in the acquisitions}
    \label{t:mca}
\end{table}

Three GAGG:Ce samples were investigated, two identical HERMES flight crystals (Sample 1 and 2), produced by C\&A (Japan), and another, differently shaped crystal (Sample 3), produced by Advatech, Inc. (UK).
Table~\ref{t:samples} summarizes the physical properties of the investigated samples.

\begin{table}[htbp]
    \centering
    \begin{tabular}{c|c|c}
        Sample & Dimensions & Producer\\ \hline
        1 & $6.94 \times 12.10 \times 15.0$  mm$^3$ & C\&A \\
        2 & $6.94 \times 12.10 \times 15.0$  mm$^3$ & C\&A \\
        3 & $10\times 10 \times 30$ mm$^3$ & Advatech \\ 
    \end{tabular}
    \caption{The GAGG:Ce samples under analysis and their physical dimensions}
    \label{t:samples}
\end{table}

\begin{figure}
    \centering
    \includegraphics[width=\textwidth]{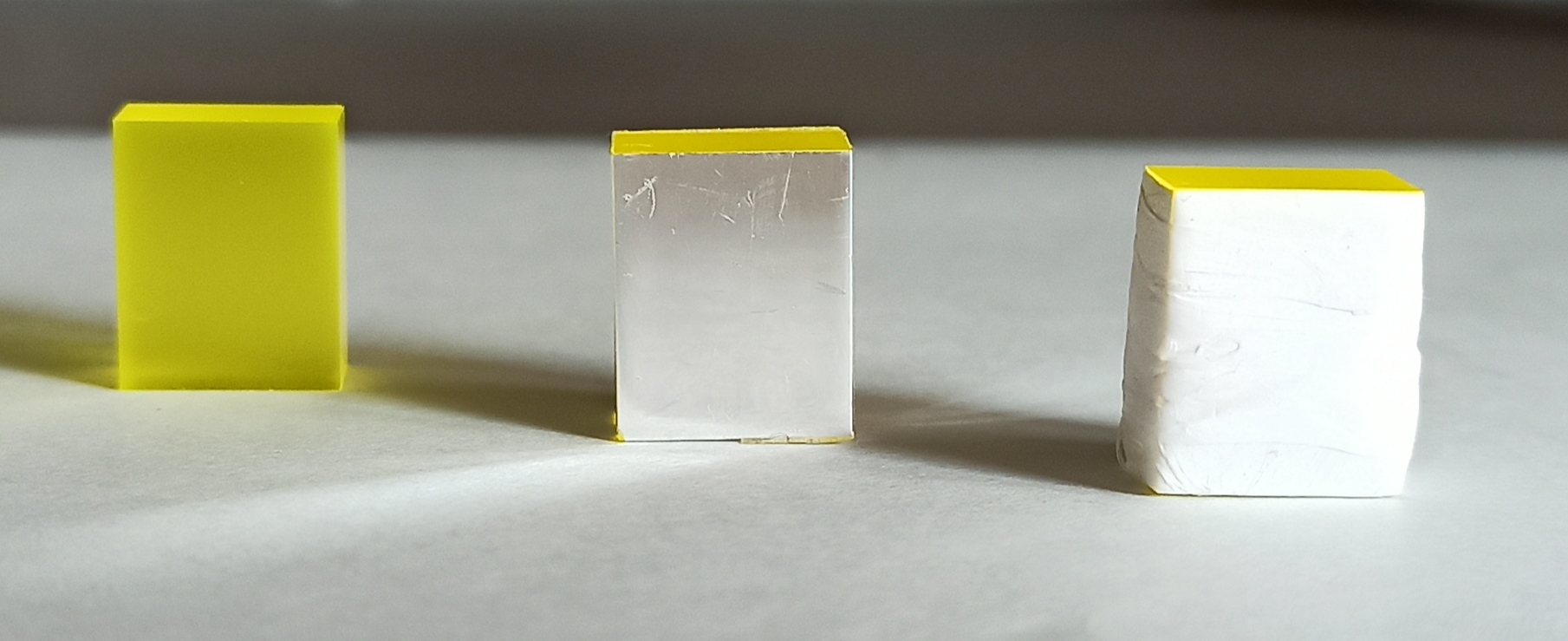}
    \caption{The HERMES $6.94\times12.1\times15$~mm$^3$ flight scintillator (Samples 1 and 2). From left to right: the crystal, the crystal with ESR, the crystal with ESR and Teflon tape.}
    \label{fig:scint_wrap}
\end{figure}

\begin{figure}
    \centering
    \includegraphics[width=0.75\textwidth]{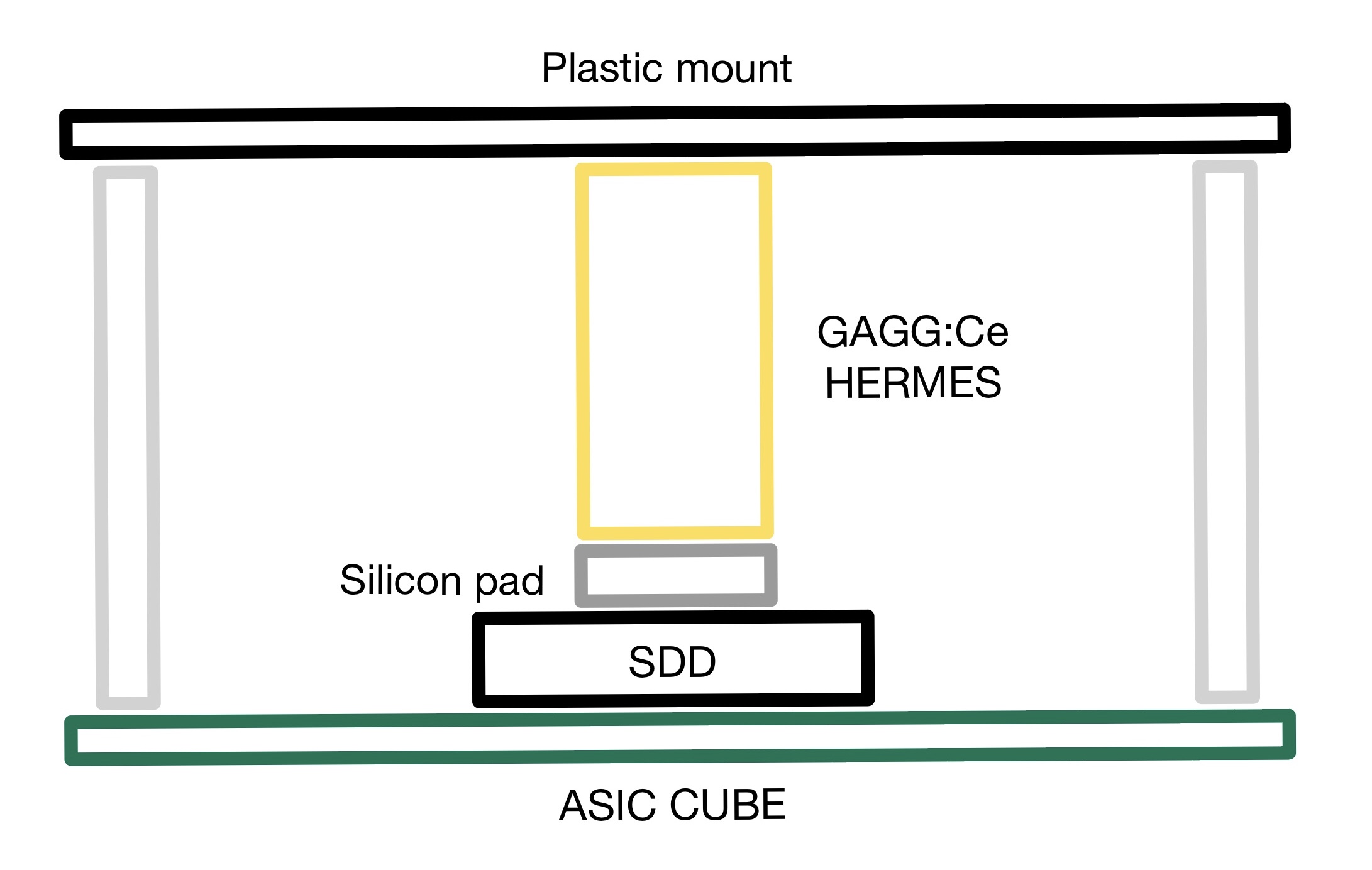}
    \caption{Sketch of the detector setup}
    \label{fig:detector_setup}
\end{figure}

Each crystal was wrapped with an enhanced specular reflector (ESR, by 3M$^{\tiny{\circledR}}$) and with a Teflon tape to minimise the loss of light at the optical surfaces (Figure~\ref{fig:scint_wrap}), then coupled to the SDD by means of a soft silicone sheet (Dowsil 93-500), about 1 mm thick. A plastic holder ensures the mechanical coupling between crystal and SDD, by pressing the crystal towards the photodetector. The detector assembly, sketched in Figure~\ref{fig:detector_setup}, is then placed inside a 2~mm thick aluminium box, to ensure electromagnetic interference and ambient light shielding. On the top side of the box a  small entrance window for X-rays has been placed, covered with aluminium tape.

\subsection{The LARIX-A facility}

\begin{figure}
    \centering
    \includegraphics[width=1\textwidth]{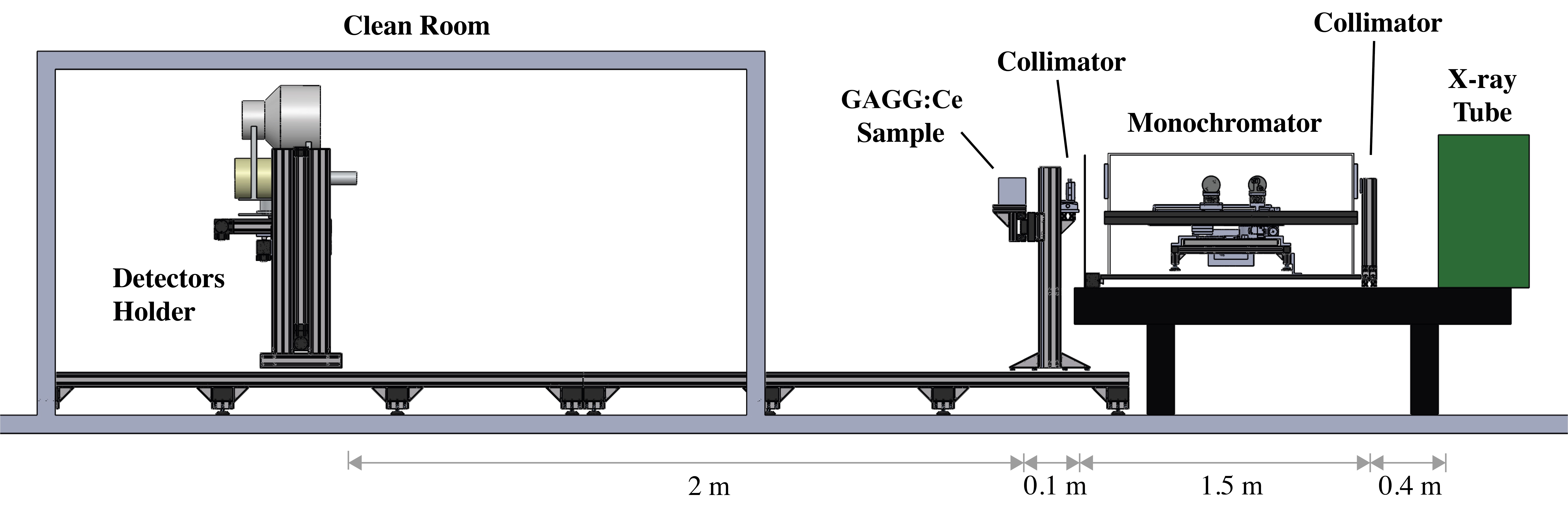}
    \caption{Sketch of the LARIX-A facility setup. From right to left: X-ray tubes, monochromator in the plexiglass box, collimator, sample holder and HPGe detector in the clean room.}
    \label{fig:larx-a}
\end{figure}

The LARIX (LARge Italian X-ray facility) laboratory\footnote{\url{https://larixfacility.unife.it/}} is located in the Scientific and Technological Pole of the University of Ferrara and is a multi-project facility. It includes LARIX-A, a 12~m long X-ray facility equipped with a Bosello X-ray tube coupled with a fixed-exit Bragg-Bragg monochromator (Figure~\ref{fig:larx-a}) providing a monochromatic beam from about 10~keV to 200~keV. The X-ray tube features a tungsten anode and provides a power of 1.8~kW with voltages up to 225~kV, and the monochromator features two mosaic Si(111) crystals with size $80\times40\times2$~mm$^3$ and mosaicity of 30$''$ to extract a single energy from the broadband X-ray tube output spectrum and redirect it along a direction parallel to the incident beam. To reduce the X-ray absorption in air, the monochromator is placed inside a plexiglass vacuum tight box that can be filled with helium. Likewise, a vacuum tube can be placed along the beamline. The software to control the monochromator by means of micrometric actuators is implemented in National Instrument's LabVIEW\texttrademark. The facility also includes two beam monitoring detectors inside a class $10^5$ clean room: an ORTEC nitrogen-cooled HPGe spectrometer (549~eV at 122~keV) with a beryllium entrance window; and a THALES X-ray imager detector with a CsI scintillator X-ray detection unit coupled with a CCD camera. A pair of collimators, one before and one after the monochromator, control the beam divergence and size. 

\begin{figure}
    \centering
    \includegraphics[width=0.75\textwidth]{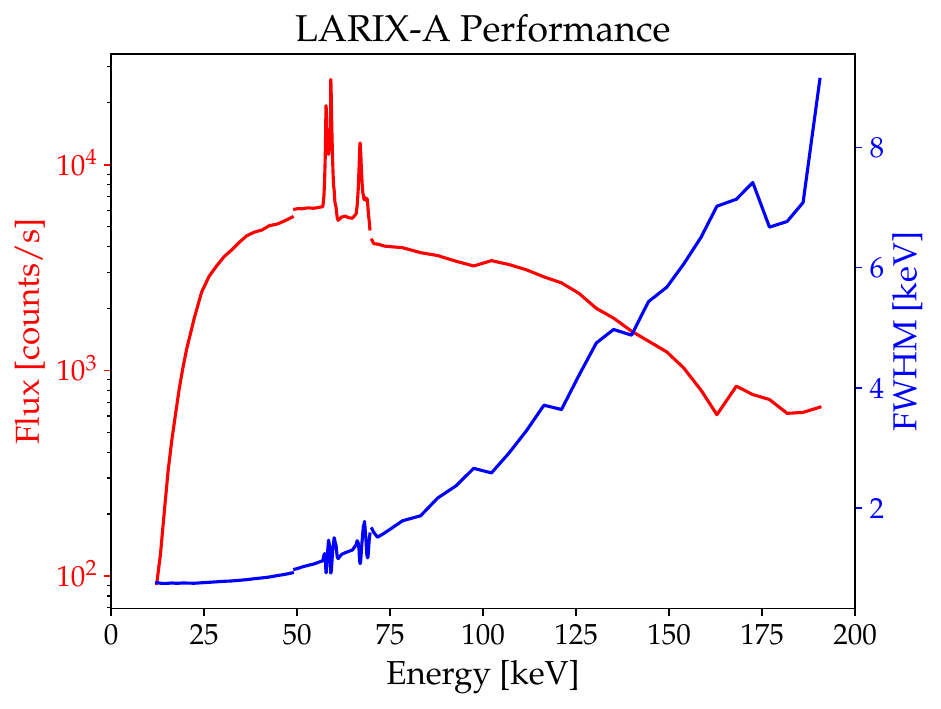}
    \caption{LARIX-A monochromatic beam performance vs. energy. The red curve reports the beam count rate, while the blue curve the full-width at half maximum of the monochromatic beam.}
    \label{fig:larix_performance}
\end{figure}

Before the experiment, the performance of the monochromatic beam as a function of the energy was evaluated using the HPGe spectrometer placed about 3~m from the position of the sample (Figure~\ref{fig:larix_performance}). The flux observed has a direct relation with the polychromatic beam generated by the X-ray tube, decreasing at higher energies, typical features of Bremsstrahlung radiation, and show two peaks at 58~keV and 67~keV, corresponding to the $K_\alpha$ and $K_\beta$ W lines, respectively. The decrease at lower energies is also due to the absorption of the low energy photons along the beamline. The measurements reported in this paper were all performed in air.

\section{Measurements}
\label{sec:meas}
\begin{figure}
    \centering
    \includegraphics[width=0.75\textwidth]{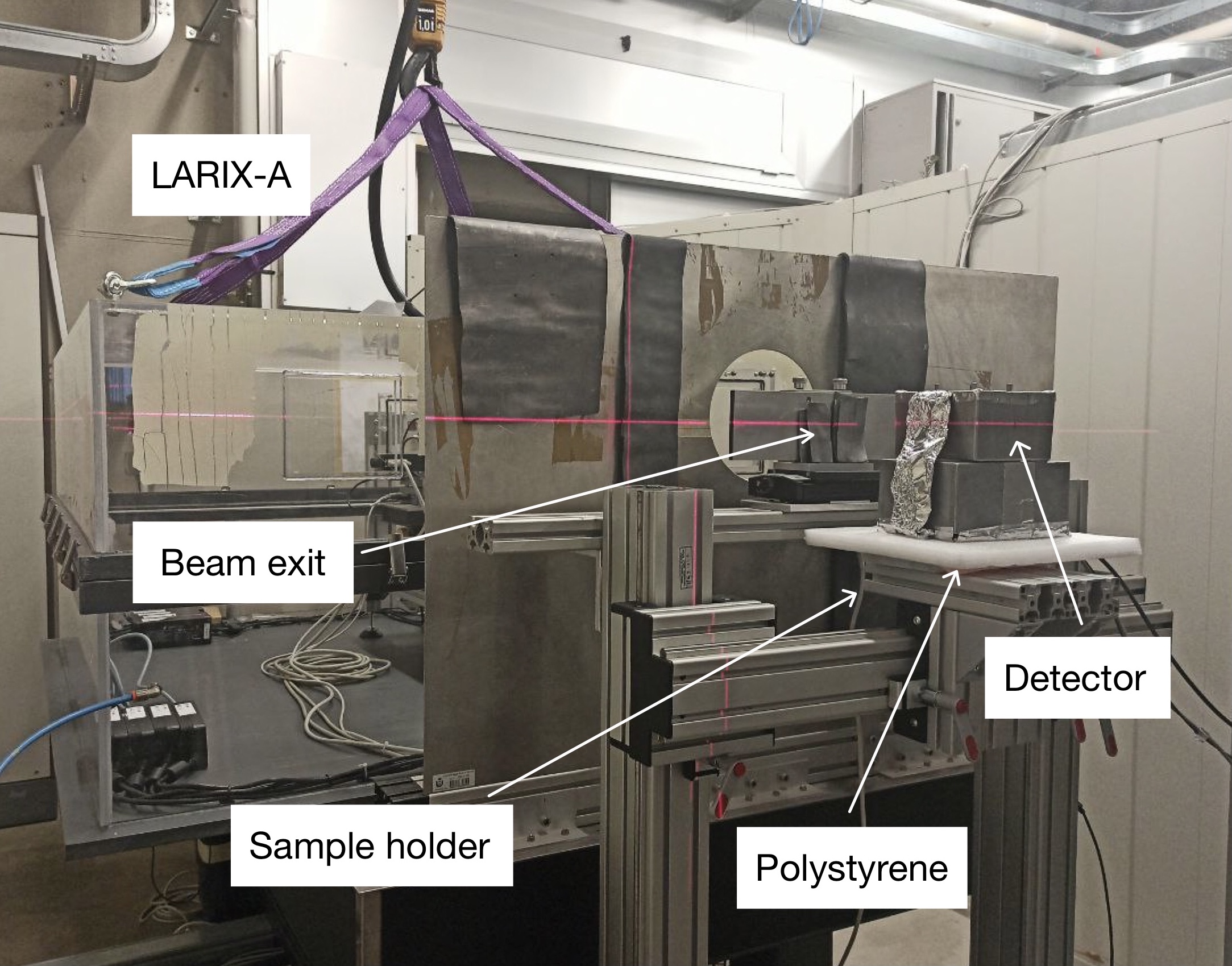}
    \caption{The detector setup mounted at the LARIX facility.}
    \label{fig:detector_setup_at_larix}
\end{figure}

The experimental setup is shown in Figure~\ref{fig:detector_setup_at_larix}. The detector box is mounted on a sample holder (electrically insulated from the holder by means of a polystyrene interposer, to minimize electromagnetic interferences), aligned to the beam exit at a distance of about 10~cm from the second collimator. The beam spot size on the crystal is estimated to be about $1\times1$ mm$^2$, and is therefore completely contained within the crystal cross-section.

The measurement campaigns on each sample followed, with little variations, these main steps:
\begin{enumerate}
    \item Calibration of the SDD
    \item Acquisition of different radioactive source spectra
    \item Acquisition of beam spectra at different nominal energies
\end{enumerate}
As will be discussed in more detail below, the first step allows to determine the absolute charge collected at the SDD anode, which is proportional to the amount of collected scintillation light, and therefore to the intrinsic light yield. While the monochromatic beam spans the interval between $\sim$20~keV to $\sim$150~keV, the use of radioisotopes allows both to expand the investigated range up to 662 keV, but also to monitor and correct the calibration of the nominal beam energy itself.

Measurements on each sample were performed on different days, with a rather stable ambient temperature around $22\pm1$~$^\circ$C. 

\subsection{Detector calibration}\label{s:det_calib}
To calibrate the SDD we can exploit the fact that it is sensitive to both X-rays and optical scintillation light. For each high-energy X-ray directly absorbed in the Si bulk, on average one electron-hole pair is produced for every $\sim$3.65~eV of deposited energy \cite{mazziotta08}. As a consequence, by illuminating the SDD with a $^{241}$Am radioactive source, which produces several X and $\gamma$-ray lines at energies between 11 and 60~keV, it is possible  to associate to each MCA digital channel (expressed in ADC units) the corresponding signal charge amplitude at the detector anode.
The calibration with $^{241}$Am was repeated several times at the beginning, in the middle and at the end of each beam data acquisition, to monitor possible gain/offset drifts (e.g., temperature-dependent). 

\begin{figure}
    \centering
    \includegraphics[width=0.75\textwidth]{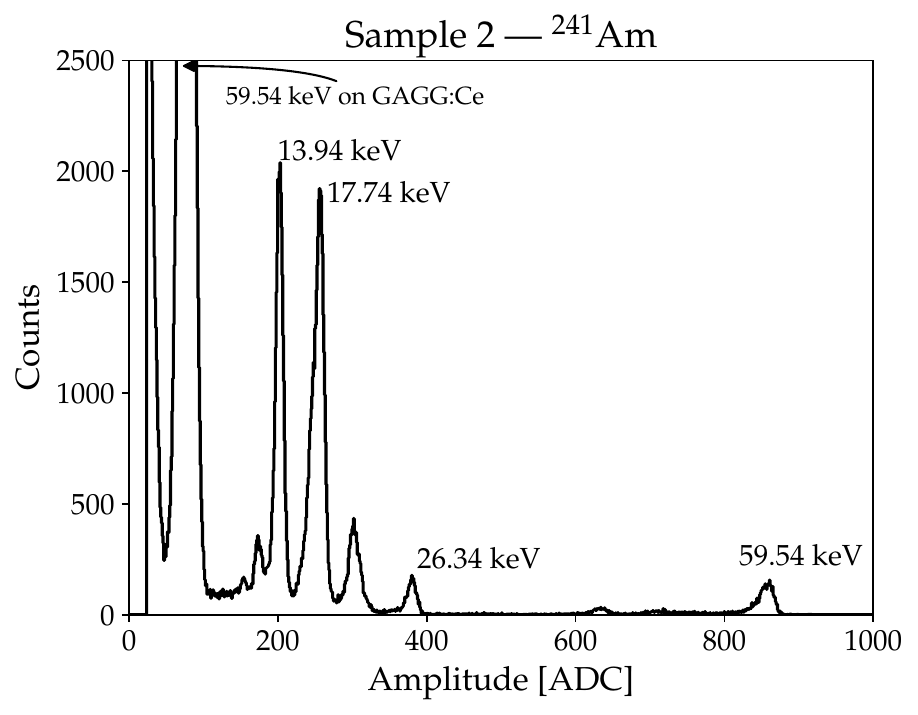}
    \caption{Spectrum of a $^{241}$Am source illuminating the SDD. The integration time is 300~s.The main X-ray lines are indicated. The broad and intense peak at low amplitudes is due to the scintillation light produced when the 59.54~keV $\gamma$-ray is absorbed in the crystal (in this case sample 2, but results are completely similar for the other samples), instead of interacting in the Si bulk of the detector. The small peak around $\sim$620 ADC is due to the fluorescence of the Gd ($K_\alpha$ line at 42.98~keV), excited by the 59.54~keV $\gamma$-rays and escaping from the crystal.}
    \label{fig:am241}
\end{figure}

A $^{241}$Am calibration spectrum is shown in Figure~\ref{fig:am241}. The positions of the peaks at 13.94, 17.74, 26.34, and 59.54 keV (corresponding to a charge of 3872, 4928, 7317, and 16538 e$^-$, respectively) were determined by means of a Gaussian fit, and then a linear interpolation allows to determine the calibration relation.

\subsection{Radioactive source acquisition}
For each sample, spectra were acquired for several radioactive sources (see Table~\ref{t:sources}), emitting monochromatic photons over a large energy interval. In each case, the source was placed on the side of the detector box, to maximise the exposed area of the scintillator crystal and to minimise the direct absorption of X-rays in the SDD. The integration time for each measurement was chosen between 5 and 10 minutes, depending on the specific source activity.

\begin{table}[htbp]
    \centering
    \begin{tabular}{c|c}
        Isotope & Energy [keV] \\ \hline
       $^{241}$Am & 59.54  \\ 
        (9.5 $\mu$Ci)     &        \\ \hline
       $^{137}$Cs   & 32.88*  \\
         (5.1 $\mu$Ci)           & 661.65 \\ \hline
        $^{57}$Co   & 122.06 \\ 
       (0.10 $\mu$Ci)     &        \\\hline
        $^{22}$Na   & 511.00 \\ 
        (0.18 $\mu$Ci)    &        \\\hline
        $^{133}$Ba  & 80.89*  \\  
          (1.5 $\mu$Ci)   & 276.39 \\
                    & 302.85 \\
                    & 356.02 \\ 
    \end{tabular}
    \\
    \footnotesize{* Weighted average over a line multiplet.}
    \caption{The radioactive sources used in the measurements and their typical photon energies. Source activities are as calculated at the time of the measurements.}
    \label{t:sources}
\end{table}

\subsection{Monochromatic beam}
For each sample, spectra with an integration time of 60 seconds were acquired from the lowest detectable energy of 20--40 keV up to a maximum value of 130--160 keV, above which the broadening of the beam FWHM (cf. Figure~\ref{fig:larix_performance}) makes difficult to distinguish the main line. The energy range 48--52~keV, around the 50.23~keV Gd $K$-edge, was carefully studied by acquiring spectra at 0.25~keV intervals, while, to study the generic light output, up to 100 keV spectra were acquired with 2~keV steps and above 100 keV at 5--10~keV steps. During the measurements, the voltage and current settings of the X-ray tube were adjusted according to the energy, in order to minimise the occurrence of pile-up. In any case, some of the spectra show residual pile-up phenomena that, however, do not  affect the subsequent analysis, as discussed in the following.

\section{Data analysis and results}
\label{sec:res}
The spectra acquired with radioactive sources and monochromatic beam were analyzed by identifying the relevant emission lines. 

\begin{figure}
    \centering
    \includegraphics[width=0.75\textwidth]{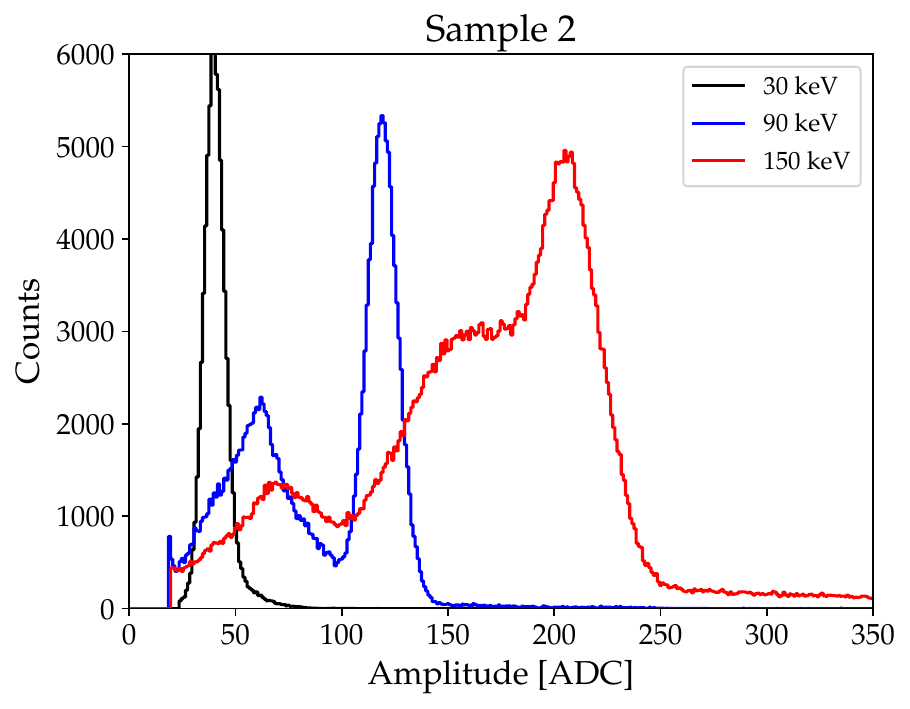}
    \caption{Example of three monochromatic beam acquisitions on sample 2 at nominal beam energies of 30, 90 and 150~keV. At higher energies the backscattering bump and the Compton continuum are also evident. The integration time is 60~s for each measurement.}
    \label{fig:beam_examples}
\end{figure}

The spectra acquired with the monochromatic beam show different profiles. Some examples are shown in  Figure~\ref{fig:beam_examples}: at lower energies the spectrum is described by a single Gaussian line, while at increasing energies the spectrum shows, in addition, the Compton continuum and the characteristic backscattering bump. Each acquisition has been therefore modelled using an appropriate phenomenological model composed by one or more Gaussian functions.
Moreover, in a few acquisitions, due to pile-up, a low-intensity and broad secondary peak or bump centred at twice the amplitude of the main monochromatic photopeak is present. This secondary bump usually does not affect the modelling, since it is sufficiently far from the main peak to be safely excluded from the analysis, or it can be anyway modelled by adding a further Gaussian function to the fit.

The spectral lines produced by the radioactive sources were similarly modelled with Gaussian functions.

Each fitted line amplitude (in units of the multichannel analyzer ADC) is then converted to electrons, using the $^{241}$Am calibration data (Section~\ref{s:det_calib}).

The nominal beam energy, corresponding to each amplitude, is then corrected for possible drifts due to the temperature or mechanical variations of the Bragg-Bragg monochromator, by comparing the positions of the scintillation peaks measured with the various radioactive sources, almost-simultaneously acquired (e.g., the 59.54~keV $^{241}$Am or the 122.06~keV $^{57}$Co line) with a linear interpolation of the nearest beam acquisitions. Deviations up to a 1\% in the nominal beam energy were measured, and thus corrected.

\begin{figure*}
    \centering
    \includegraphics[width=\textwidth]{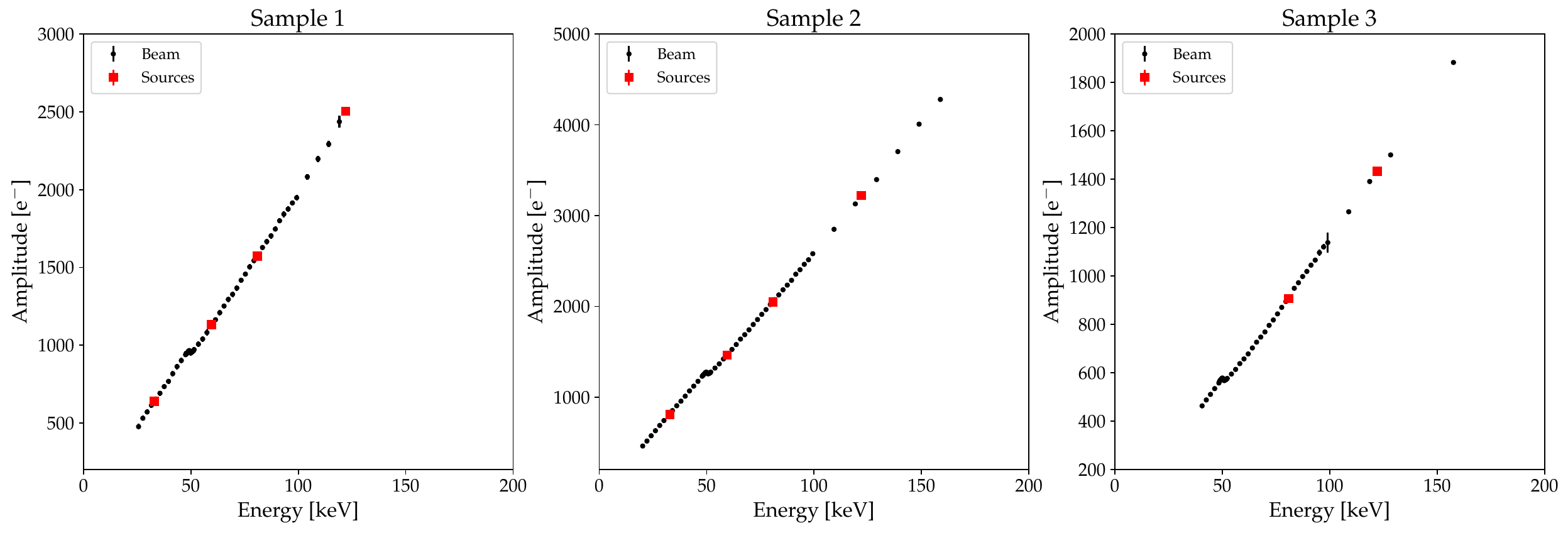}
        \includegraphics[width=\textwidth]{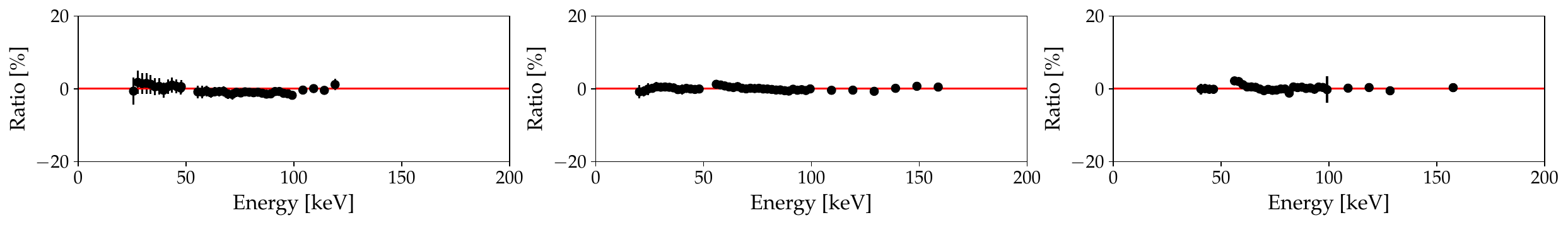}
    \caption{Amplitude of the signal, in electrons, versus energy for the three samples. The lower panel shows the residuals with respect to a linear fit below and above 50.23~keV.}
    \label{fig:amplitudes_vs_energy}
\end{figure*}

The energy-corrected peak amplitudes for the beam are shown in Figure~\ref{fig:amplitudes_vs_energy}. A discontinuity around the 50.23~keV Gd $K$-edge is well apparent, with a broadly linear behavior both above and below this energy (Figure~\ref{fig:amplitudes_vs_energy}, lower panel).

The signal amplitude in electrons $N_{e^-}$, as a function of the photon energy $E_\gamma$, is then modelled as a linear function:
\begin{equation}
    N_{e^-} = a E_\gamma + b
\end{equation}
The effective light output $LY_\mathrm{eff}$ (which is proportional to the intrinsic light yield $LY$ through the optical contact efficiency and the photodetector quantum efficiency) is defined as:
\begin{equation}
    LY_\mathrm{eff} = \frac{N_{e^-}}{E_\gamma}
\end{equation}
and thus:
\begin{equation}\label{eq:model}
    LY_\mathrm{eff} = a + \frac{b}{E_\gamma}
\end{equation}

\begin{figure*}
    \centering
    \includegraphics[width=\textwidth]{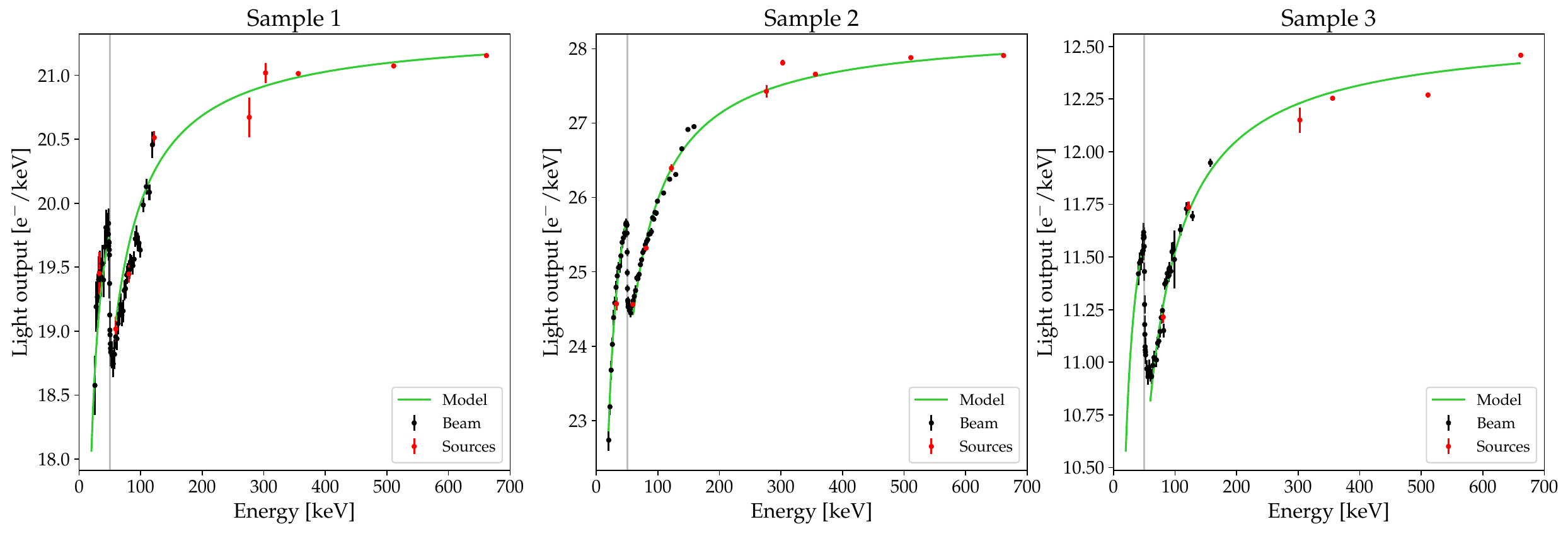}
    \caption{The measured light outputs vs energy, with their fit using the model in Eq.~\ref{eq:model} for the three samples. Since the 276~keV $^{133}$Ba line for Sample 3 could not be fitted well, it was omitted in the analysis.}
    \label{fig:amplitudes_vs_energy_fitted}
\end{figure*}

The modelling is performed separately on the two sides of the $K$-edge discontinuity, below $\sim$48~keV and above $\sim$52 keV, respectively, for each sample. The fitted parameters are reported in Table~\ref{t:fits} and shown in the three panels of Figure~\ref{fig:amplitudes_vs_energy_fitted}.

\begin{table}[htbp]
    \centering
    \begin{tabular}{c|c|c|c|c|c}
        Sample & $a$ & $b$ & $a'$ & $b'$ & LY$_\mathrm{eff}$(662~keV)\\ 
               & [e-/keV]  & [e$^-$] & [e$^-$/keV]  & [e$^-$] & [e-/keV] \\ 
        \hline
        1 & $20.92\pm0.16$ & $-57.1\pm6.5$ & $21.37\pm0.02$ & $-136.8\pm2.8$ & $21.15 \pm 0.01$ \\
        2 & $27.64\pm0.07$ & $-95.5\pm2.5$ & $28.28\pm0.02$ & $-231.0\pm3.0$ & $27.91 \pm 0.01$ \\
        3 & $12.25\pm0.14$ & $-33.4\pm6.3$ & $12.58\pm0.02$ & $-105.7\pm3.0$ & $12.46 \pm 0.01$ \\
    \end{tabular}
    \caption{Fitted values for the modelling. $a$ and $b$ are measured below $\sim$48~keV, while $a'$ and $b'$ above $\sim$52~keV. The last column reports the measured effective light output at the $^{137}$Cs 662~keV energy.}
    \label{t:fits}
\end{table}

\begin{figure}
    \centering
    \includegraphics[width=0.75\textwidth]{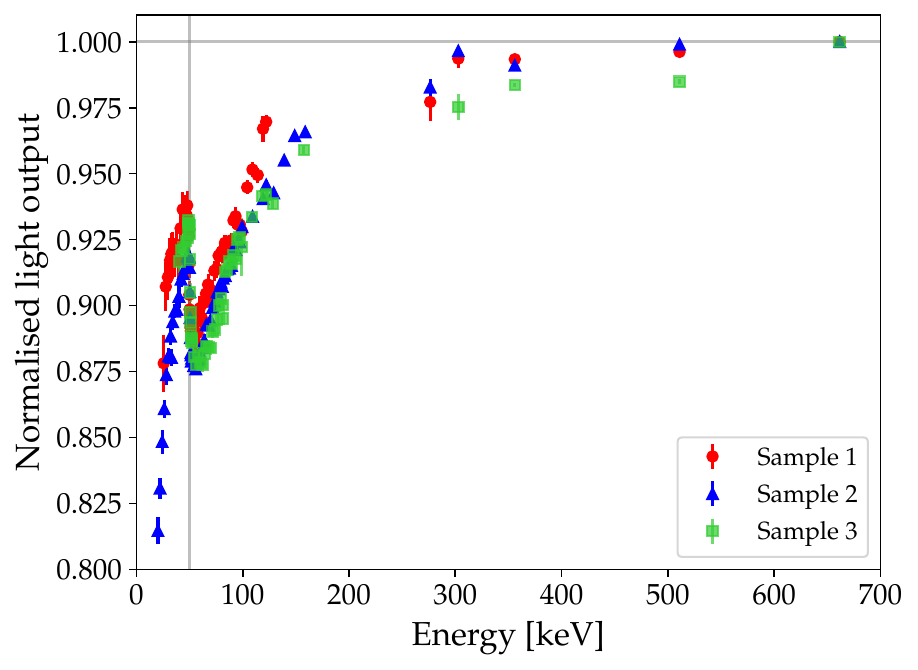}
    \caption{Light output as a function of the energy, normalised at its 662 keV value, for the three GAGG:Ce samples under investigation. The vertical line marks the 50.23~keV Gd $K$-edge.}
    \label{fig:final_ly}
\end{figure}

Figure~\ref{fig:final_ly} shows the measured light outputs, normalised to their 662~keV value taken as a reference. The three samples, as reported in Table~\ref{t:fits}, have different values of the absolute light output (depending on the crystal geometry and/or optical coupling efficiency between scintillator and SDD), but show a rather similar relative behavior, with a continuous decrease of the light yield below $\sim$200~keV of a few tens of percent down to $20$~keV, superimposed to a 4\%--5\% ``jump'' around the Gd $K$-edge.

\section{Discussion and conclusions}
\label{sec:disc}
The measurement of the GAGG:Ce scintillator non-linearity (or, more appropriately, the decrease in the specific light output with respect to the reference value measured at 662~keV) is of paramount importance for a correct calibration of the HERMES detectors \cite{campana22}. Using the monochromatic photon beam produced at the LARIX facility, and several known lines from radioactive isotopes, the overall light output of the crystal has been measured on three samples, two HERMES flight crystals and, for the sake of comparison, on another crystal by a different producer and with a different geometry. The results, summarized in Figure~\ref{fig:final_ly}, are broadly consistent with previous measurements (e.g., \cite{sibczynski18,gao22}), with a $\sim$13\% decrease in light output down to 50~keV, and an abrupt 4\%--5\% discontinuity in correspondence of the binding energy of the $K$-shell in gadolinium at 50.23~keV. 

\section*{Acknowledgements}
This work has been carried out in the framework of the HERMES-TP and HERMES-SP collaborations. We acknowledge support from the European Union Horizon 2020 Research and Innovation Framework Programme under grant agreement HERMES-Scientific Pathfinder n. 821896 and from ASI-INAF Accordo Attuativo HERMES Technologic Pathfinder n. 2018-10-H.1-2020.
This work is partly supported by the AHEAD-2020 Project grant agreement 871158
of the European Union’s Horizon 2020 Programme. 
We are grateful to the anonymous referees, whose useful comments greatly improved this paper.

 \bibliographystyle{elsarticle-num} 
 \bibliography{cas-refs}





\end{document}